\documentstyle[11pt,aaspp]{article}

\begin{document}

%
%
%
   \def\prd#1#2#3#4{#4 19#3 Phys.~Rev.~D,\/ #1, #2 }
   \def\pl#1#2#3#4{#4 19#3 Phys.~Lett.,\/ #1, #2 }
   \def\prl#1#2#3#4{#4 19#3 Phys.~Rev.~Lett.,\/ #1, #2 }
   \def\pr#1#2#3#4{#4 19#3 Phys.~Rev.,\/ #1, #2 }
   \def\prep#1#2#3#4{#4 19#3 Phys.~Rep.,\/ #1, #2 }
   \def\pfl#1#2#3#4{#4 19#3 Phys.~Fluids,\/ #1, #2 }
   \def\pps#1#2#3#4{#4 19#3 Proc.~Phys.~Soc.,\/ #1, #2 }
   \def\nucl#1#2#3#4{#4 19#3 Nucl.~Phys.,\/ #1, #2 }
   \def\mpl#1#2#3#4{#4 19#3 Mod.~Phys.~Lett.,\/ #1, #2 }
   \def\apj#1#2#3#4{#4 19#3 Ap.~J.,\/ #1, #2 }
   \def\aj#1#2#3#4{#4 19#3 Astr.~J.,\/ #1, #2}
   \def\acta#1#2#3#4{#4 19#3 Acta ~Astr.,\/ #1, #2}
   \def\rev#1#2#3#4{#4 19#3 Rev.~Mod.~Phys.,\/ #1, #2 }
   \def\nuovo#1#2#3#4{#4 19#3 Nuovo~Cimento~C,\/ #1, #2 }
   \def\jetp#1#2#3#4{#4 19#3 Sov.~Phys.~JETP,\/ #1, #2 }
   \def\sovast#1#2#3#4{#4 19#3 Sov.~Ast.~AJ,\/ #1, #2 }
   \def\pasj#1#2#3#4{#4 19#3 Pub.~Ast.~Soc.~Japan,\/ #1, #2 }
   \def\pasp#1#2#3#4{#4 19#3 Pub.~Ast.~Soc.~Pacific,\/ #1, #2 }
   \def\annphy#1#2#3#4{#4 19#3 Ann. Phys. (NY), \/ #1, #2 }
   \def\yad#1#2#3#4{#4 19#3 Yad. Fiz.,\/ #1, #2 }
   \def\sjnp#1#2#3#4{#4 19#3 Sov. J. Nucl. Phys.,\/ #1, #2 }
   \def\astap#1#2#3#4{#4 19#3 Ast. Ap.,\/ #1, #2 }
   \def\anrevaa#1#2#3#4{#4 19#3 Ann. Rev. Astr. Ap.,\/ #1, #2
                       }
   \def\mnras#1#2#3#4{#4 19#3 M.N.R.A.S.,\/ #1, #2 }
   \def\jdphysics#1#2#3#4{#4 19#3 J. de Physique,\/ #1,#2 }
   \def\jqsrt#1#2#3#4{#4 19#3 J. Quant. Spec. Rad. Transfer,\/ #1,#2 }
   \def\jetpl#1#2#3#4{#4 19#3 J.E.T.P. Lett.,\/ #1,#2 }
   \def\apjl#1#2#3#4{#4 19#3 Ap. J. (Letters).,\/ #1,#2 }
   \def\apjs#1#2#3#4{#4 19#3 Ap. J. (Supp.).,\/ #1,#2 }
   \def\apl#1#2#3#4{#4 19#3 Ap. Lett.,\/ #1,#2 }
   \def\astss#1#2#3#4{#4 19#3 Ap. Sp. Sci.,\/ #1,#2 }
   \def\nature#1#2#3#4{#4 19#3 Nature,\/ #1,#2 }
   \def\spscirev#1#2#3#4{#4 19#3 Sp. Sci. Rev.,\/ #1,#2 }
   \def\advspres#1#2#3#4{#4 19#3 Adv. Sp. Res.,\/ #1,#2 }
   %
%
%
\def\Msun{M_{\odot}}
\def\Mdot{\dot M}
\def\deg{$^\circ$\ }
\def\etal{{\it et~al.\ }}
\def\eg{{\it e.g.,\ }}
\def\etc{{\it etc.}}
\def\ie{{\it i.e.,}\ }
\def\ksec{{km~s$^{-1}$}}
\def\arcsec{{$^{\prime\prime}$}}
\def\arcmin{{$^{\prime}$}}
\def\subsun{_{\twelvesy\odot}}
\def\sun{\twelvesy\odot}
\def\gtwid{\mathrel{\raise.3ex\hbox{$>$\kern-.75em\lower1ex\hbox{$\sim$}}}}
\def\ltwid{\mathrel{\raise.3ex\hbox{$<$\kern-.75em\lower1ex\hbox{$\sim$}}}}
\def\plusminus{\mathrel{\raise.3ex\hbox{$+$\kern-.75em\lower1ex\hbox{$-$}}}}
\def\minusplus{\mathrel{\raise.3ex\hbox{$-$\kern-.75em\lower1ex\hbox{$+$}}}}

\title{The Quasi-Stationary Structure of Radiating Shock Waves.\\
        II. The Two-Temperature Fluid}

\author{M W Sincell}
\affil{Département d'Astrophysique Extragalactique et de Cosmologie\\
 Observatoire de Paris-Meudon}
\authoraddr{Département d'Astrophysique Extragalactique et de Cosmologie\\
 Observatoire de Paris-Meudon\\ 92125 Meudon Cedex \\ FRANCE}

\author{M Gehmeyr, D Mihalas}
\affil{Department of Astronomy, University of Illinois at Urbana-Champaign
\\Laboratory for Computational Astrophysics, National Center for
Supercomputing
Applications}
\authoraddr{Department of Astronomy \\University of Illinois at
Urbana-Champaign \\1110 W. Green Street \\Urbana, IL 61801-3080}

\begin{abstract}

We solve the equations of radiation hydrodynamics in the two-temperature
fluid approximation on an adaptive grid.  The temperature structure depends
upon the electron-ion energy exchange length, $l_{ei}$,
and the electron conduction length, $l_{ec}$.  Three types of radiating shock
structure are observed: subcritical, where preheating of the unshocked gas
is negligible; electron supercritical, where radiation preheating raises the
temperature of the unshocked electron fluid to be equal to the final electron
temperature; supercritical, where preheating and electron-ion energy
exchange raise the preshock $T_{e,i}$  to their final post shock values.
No supercritical shock develops when $l_{ei}$ is larger than the photospheric 
depth of the shocked gas because a negligible amount of the ion energy is 
transferred
to the electrons and the shock is weakly radiating. Electron conduction
smooths the $T_e$ profile on a length scale $l_{ec}$, reducing the
radiation flux.

\end{abstract}

Keywords:  Radiating shock waves -- Numerical methods

Correspondence to:  M. W. Sincell, Mark.Sincell@daec.obspm.fr

\section {Introduction} 
\label{sec: introduction} 

The structure and dynamics of radiative shock waves are difficult to 
model because processes in the shock front occur
on length scales that are many orders of magnitude smaller than the
typical length scales for other gradients in the fluid variables (\eg the
velocity field in an accretion flow).  There are two standard methods for
computing the structure of shocked fluids.  The first is to treat the
shock as a discontinuity and invoke conservation laws to relate physical
quantities on either side of the shock.  Analytic models of shock waves in
plasmas have been constructed using this approach (Zel'dovich and
Raizer 1967, Shafranov 1967)  but these solutions require many simplifying
assumptions which limit the applicability of the results.  The second
method, common in numerical solutions, is to introduce an expression for an 
artificial viscosity to spread the shock over a few grid points. The 
magnitude of the artificial viscosity
is usually chosen to be many orders of magnitude larger than the physical
viscosity because the Courant limit imposes strong constraints on the
maximum time step (Klein, \etal 1983, Burger and Katz 1983).
                                                         
Formulating the numerical problem on an adaptive grid can dramatically
increase the effective resolution of the grid and reduce the spurious
effects of artificial viscosity.  Dorfi and Drury (1987) solved the
one-dimensional hydrodynamic equations on an adaptive grid.  They adopted
a simple grid equation which distributes grid points uniformly along the
arc length of a graph of the solution variables and solved two standard
problems: the shock tube and a spherical blast wave.  In both cases, the
adaptive grid concentrated many grid points at discontinuities in the
flow.  Although artificial viscosity is still needed to spread the
discontinuity over a few grid points, the physical separation of each
point is small compared to the length scale of the gradients in the physical
quantities and the shock front appears infinitely steep.  Gehmeyr
and Mihalas (1994) demonstrated that this same equation can be used
to resolve discontinuities in radiating flows and they performed a
preliminary numerical study of radiating shock waves.  A detailed study
of the structure of a radiating shock wave for a single temperature
fluid was carried out in Sincell, \etal (1997).
In this paper, we extend the work of Sincell, \etal (1997) to a fully 
ionized plasma.

The gas upstream of the shock is assumed to be cold and at rest.  However,
we assume that the gas is always fully ionized. A supersonic piston (speed
$u_p$)  drives a collisional shock wave into the cold gas and the wave 
propagates into
the unshocked material at a speed $D>u_p$. The structure of the shock
front is steady when viewed in a reference frame moving with the front
and, in this frame, the upstream gas flows into the shock at the shock
speed $D$.  The shocked gas moves away from the discontinuity at a
velocity $D - u_p$. 

The kinetic energy of the inflowing gas is converted into thermal energy
of the ions.  The ratio of the kinetic energy transferred to the ions to
that transferred to electrons is $\sim m_i / m_e$, where $m_{i,e}$ are the
masses of the ions and electrons, respectively.  As a consequence, the
increase in the electron temperature caused by viscous heating at the
shock front is negligible.  The dominant source of electron heating at the
shock front is adiabatic compression. The plasma remains neutral and so
the electron number density must change in strict proportion with the ion
density.  This results in compressional heating of the electrons as the
gas passes through the discontinuity. For a gas with an adiabatic index of
5/3, this increases the electron temperature by at most a factor of 2.5
(Zeldovich and Raizer 1967).

The ratio of the electron and ion temperatures outside of the shock front
is determined by two length scales.  The ion temperature exceeds the
electron temperature to a distance $l_{ei} \sim \tau_{ei} D$ behind the
shock, where $\tau_{ei}$ is the time scale for energy exchange between the
electron and ion fluids.  Electron conduction transports energy over a
distance $l_{ec} \sim \kappa_{ec} / D$, where $\kappa_{ec}$ is the
electron conduction coefficient.  Conduction can raise the preshock
electron temperature above the ion temperature. 

The electron gas upstream from the shock is also heated by radiation from
the shocked gas.  If the shock is strong enough, the temperature of the
preheated electron fluid rises to be equal to the temperature of the
shocked gas.  At this strength the shock is called supercritical.  A full
discussion of sub- and supercritical shock waves is
found in Sincell, \etal (1997). 

We compute the structure of a radiating shock wave in a fully ionized gas
for a simple model problem: a piston moving supersonically through a
spherical shell of cold gas at initially constant density.  We also assume
that the electron-ion energy exchange rate is directly proportional to the
difference in the electron and ion temperatures and the electron
conduction flux is proportional to the electron temperature gradient.  The
proportionality coefficients are all taken as constants.  Although this
model is too simplified to treat realistic problems, it demonstrates the
power of the adaptive grid when applied to two-temperature flows and
illustrates the effects of conduction and electron-ion energy exchange on
the structure of the shock wave.

The equation set and methodology are discussed in chapter \ref{sec: equations
and methodology} and the results for a series of models are presented
in chapter \ref{sec: results} We conclude in chapter \ref{sec: conclusions}

\section{Equations and Methodology}
\label{sec: equations and methodology}

We use the TITAN code (Gehmeyr and Mihalas 1994, Sincell, \etal 1997) to 
solve the 
time-dependent 
equations of radiation hydrodynamics on an adaptive grid. 
Gehmeyr and Mihalas (1994) provide a detailed description of 
TITAN so we will only summarize the key features of the code here.
The equations of radiation hydrodynamics in the two fluid approximation
(electron and ions are treated as separate fluids)
are the continuity equation
\begin{equation}
\label{eq: continuity}
D_t(\rho) + \rho { \partial (r^2 u) \over r^2 \partial r} = 0,
\end{equation}
the gas momentum equation
\begin{equation}
\label{eq: gas momentum}
\rho D_t(u) + {\partial P_e \over \partial r} + {\partial (r^3 P_Q) \over
r \partial r} - {\rho \kappa \over c} F_r = 0,
\end{equation}
the radiation momentum equation
\begin{equation}
\label{eq: rad momentum}
\rho D_t( {F_r \over \rho c^2} ) + {\partial P_r \over \partial r}
+ { {3P_r - E_r}\over{r} } + {F_r \over c^2} {\partial u \over \partial r}
+ {\rho \kappa \over c} F_r = 0,
\end{equation}
the radiation energy equation
\begin{equation}
\label{eq: rad energy}
\rho D_t({E_r \over \rho}) + { \partial(r^2 F_r) \over r^2 \partial r}
+ P_r {\partial(r^2 u) \over r^2 \partial r} 
+ (E_r - 3 P_r) {u \over r} + \rho \kappa c \left(E_r - a_r T_e^4 \right)
=0,
\end{equation}
the ion energy equation
\begin{equation}
\label{eq: ion energy}
\rho D_t(e_i) + 
P_i {\partial(r^2 u) \over r^2 \partial r}
+ P_Q \left[ {\partial u \over \partial r} - {u \over r} \right]
+ \Lambda_{ei} {k_B \over (\gamma-1) m_p} (T_i - T_e) = 0;
\end{equation}
and the total energy equation
\begin{eqnarray}
\label{eq: fluid energy}
\rho D_t(e_e + e_i + {E_r \over \rho}) + { \partial(r^2 F_r) \over r^2 \partial r}
& + & (P_e + P_i + P_r) {\partial(r^2 u) \over r^2 \partial r}
+ P_Q \left[ {\partial u \over \partial r} - {u \over r} \right] 
\nonumber \\
& + & (E_r - 3 P_r) {u \over r} 
+ {\partial \over \partial r} \left( \rho \kappa_{ec} r^2 {\partial T_e \over
\partial r} \right) = 0,
\end{eqnarray}
where $D_t(x) = \partial x / \partial t + u \partial x / \partial r$ is the
Lagrangean time derivative operator.
We assume a perfect gas equation of state with an adiabatic index of 
$\gamma=5/3$, 
a constant absorptive opacity ($\kappa$), electron conduction coefficient 
($\kappa_{ec}$) and the electron-ion energy exchange coefficient
($\Lambda_{ei}$).
The radiation pressure ($P_r$) and energy density ($E_r$) are related by a
variable Eddington
factor, $f_E = P_r / E_r$.  The Eddington factors a computed with a formal 
integration of the time-independent radiative transfer equation (\eg Mihalas
and Mihalas 1984) and updated
during each time-step.
The remaining variables in these equations represent the radius ($r$), the 
gas density ($\rho$), 
gas velocity ($u$), electron and ion gas pressures ($P_{e,i}$), 
electron and ion gas energies per 
unit mass ($e_{e,i}$), 
electron temperature
($T_e$), radiation flux ($F_r$) and the artificial viscosity ($P_Q$, see paper
I).  We neglect electron viscosity because it is typically smaller than
the ion viscosity by a factor $(m_e/m_i)^{1/2}$.

The radiation hydrodynamics equation set (eqs. \ref{eq: continuity}-
\ref{eq: fluid energy})  is supplemented with the adaptive grid equation
(Gehmeyr and Mihalas 1994, Dorfi and Drury 1987).  This equation distributes 
grid points so
that discontinuities in the flow are resolved.  In this work we found the
best grid performance when we resolve the mass and the gas density. 

The full set of equations is written in finite volume form using the
adaptive mesh transport theorem (Winkler, Norman and Mihalas 1984) and
then
cast into finite difference form on a staggered grid (Gehmeyr and 
Mihalas 1994).  The
difference equations are linearized around the current solution and the
solution at the next time step is calculated with a fully-implicit
Newton-Raphson iteration. 

\subsection{The Model Problem}
\label{subsec: model problem}

We consider a thin shell of gas extending from $R_i=8.0 \times 10^6$~km to
$R_o=8.7 \times 10^6$~km.  This problem is nearly plane parallel because
$R_o-R_i \ll R_o$.
Initially, the gas is at rest with constant density
($\rho_o = 7.78 \times 10^{-10}$~g/cm$^{3}$) and a shallow temperature
profile
\begin{equation}
T(r) = 10 + 75 { {r - R_i} \over {R_o - R_i} } \mbox{K}
\end{equation}
The sound speed in the gas is $c_s \ltwid 1$~km/sec.
The gas has a constant absorptive opacity $\rho \kappa = 3.115 \times
10^{-10}$~cm$^{-1}$. These parameters are chosen to correspond to optically 
thick gas accreting onto a neutron star.
Initially, the gas and radiation are in equilibrium and
there is no net flux of radiation.
     
At time $t=0$ a piston at $R_i$ starts outward at a constant velocity,
$u_p
> c_s$,
and a shock forms ahead of the piston. The shock propagates outward at a
velocity
\begin{equation}
\label{eq: shock speed}
D = u_p / (1 - \eta_{+})
\end{equation}
where $\eta_{+} = \rho_o / \rho_{+}$ is the shock compression ratio and
$\rho_{+}$
is the gas density behind the shock.  Note that $\eta_{+} \gtwid
\eta_{1}$,
where
$\eta_{1}$ is the final compression ratio, because some additional
compression
can occur as the shocked gas cools from $T_{+}$ to $T_1$.
After a short time, $t \ll (R_o-R_i)/D$,
the shock reaches a quasi-stationary state in which the structure of the
shock front is independent of time when viewed in a frame moving at a
velocity
$D$, \ie with the shock front.  We refer to the shock as quasi-stationary 
because as it propagates to larger distances geometric dilution of the 
spherical flow will cause an intrinsic time-dependence.  In addition, the
flow appears to fluctuate slightly near the piston.  These fluctuations have 
no effect on the structure of the shock.

In this paper we assume that both the electron-ion energy exchange
coefficient and the electron conduction coefficient are constant.  In this
case, 
\begin{equation} 
\Lambda_{ei} = \tau_{ei}^{-1} 
\end{equation} 
is the
inverse of the time-scale for electron-ion energy exchange.  The electron
conduction coefficient is \begin{equation} \kappa_{ec} = {k_B \over m_p}
l_e \bar v_e \end{equation} where $l_e$ is the electron mean free path and
$\bar v_e$ is the electron thermal velocity.  The characteristic length
scale for electron-ion energy exchange is $l_{ei} \sim \tau_{ei} D$ and
that for electron conduction is $l_{ec} \sim \kappa_{ec} / D$.  We neglect
ion conduction and electron viscosity because they are comparatively small
(Zel'dovich and Raizer 1967).

\section{Results}
\label{sec: results}

Zel'dovich and Raizer (1967) defined two classes of shock wave structure
for a single temperature fluid: subcritical and supercritical.  Absorption
of radiation from the shocked gas raises the temperature of the upstream
material.  A shock is called subcritical when the temperature of the
preheated gas is smaller than the final downstream temperature.  When the
shock wave is stronger, preheating can raise the temperature of the
upstream gas to be equal to, but never larger than, the final gas
temperature (Zel'dovich and Raizer 1967). 

Extending the classification scheme of Zel'dovich and Raizer (1967), 
we identify
four types of shock front:  subcritical, supercritical, electron
subcritical and electron supercritical.  When $l_{ei}$ is small compared
to the length scales for gradients in the flow variables, including the
radiation mean free path, the electron and ion temperatures are nearly
equal at all points in the flow. We classify these shocks using the
standard notation because the temperature structure of
these shocks is the same as the corresponding one-temperature shock wave.

The ion and electron temperatures decouple when $l_{ei}$ is larger than
the length scales for other gradients in the flow variables.  For
example, the length scale for radiative energy exchange in our model
problem is $l_r = 1/\rho\kappa = 3.2 \times 10^9$~cm.  In this
case, preheating primarily affects the electron gas because there is not
sufficient time for energy to be transferred from the electron gas to the
ions.  The shock structure is electron subcritical if the
temperature of the preheated electron gas is smaller than the final
temperature of the electron gas.  When the temperature of the preheated
electron gas becomes equal to the final electron temperature, the shock is
electron supercritical.  In both cases, the $T_i \ltwid T_e$ upstream from
the shock and $T_i \gtwid T_e$ downstream.

The kinetic energy of the inflowing gas is converted to thermal
energy in the ion gas at the shock discontinuity. The temperature of the
shocked ion gas is $T_{i} \sim m_i D^2 / k$, where $m_i$ is the mass of
the ions.  
The hot ion gas cools by Coulomb collisions with the cold
electrons.  If $l_{ei}$ is smaller than the shock standoff distance and
all the kinetic energy of the infalling gas is converted into thermal
energy, then the ion and electron temperatures equalize at the final value
\begin{equation}
\label{eq: T_1}
T_1 \simeq {1 \over 2} {m_p \over k_B} \eta_{1} \left( 1-\eta_{1}
\right)^{-2}
u_p^2 \simeq 27 u_{p,5}^2 \mbox{K},
\end{equation}
where $u_{p,5}$ is the piston speed in $10^5$~cm/sec.  The final
temperature of the gas flowing through a supercritical shock is somewhat
smaller than this value because the electron gas radiates a large part of
the inflowing energy.

The ion temperature in the shocked gas remains larger than the electron
temperature if $l_{ei}$ is large.  This has two consequences.  First, the
final ion temperature is larger than $T_1$ and the final electron
temperature is smaller than $T_1$.  Second, the temperature of the
preheated gas is lower because the temperature of the shocked electron gas
is lower.  This reduces the amount of energy which is radiated by the
shocked gas.  If $l_{ei}$ is large enough (section 
\ref{subsec: critical rate}),
the shock doesn't become supercritical at any piston speed. 

In the following sections, we present results of our simulations for four
sets of the parameters $l_{ei}$ and $l_{ec}$.  They illustrate the effects of
electron-ion energy exchange and electronic heat conduction on the structure
of a radiating shock wave.  In each case we plot the results for two
piston speeds, 4 km/sec and 14 km/sec, which correspond to subcritical and
supercritical shock strengths, respectively.  

Figures \ref{fig:
subtf.4.11} - \ref{fig: supertf.2.15} each contain at least two panels:
one shows the electron (dashed line) and ion (solid line) temperatures
as a function of optical depth from the shock front ($\tau$) and
the other shows the radiation (solid line) and electron conduction
(dashed line) fluxes as a
function of $\tau$. In Figs.
\ref{fig: subtf.2.11} - \ref{fig: supertf.2.15} we also include a third
panel containing a detail of the temperature structure in the shock front.
The values of $l_{ei}$ and $l_{ec}$ for the models in this paper are listed
in table \ref{table: model parameters}.

\subsection{$\Lambda_{ei} = 10^{-4}$~sec$^{-1}$, 
        $l_e \bar v_e = 10^{11}$~cm$^2/$~sec}
\label{subsec: small small}

The temperature and radiation profiles of the subcritical shock wave are
shown in Figs. \ref{fig: subtf.4.11}a,b. This is an example of the
structure of an electron
subcritical shock wave.
The electron-ion energy exchange length is on the order of 1\% of
the total width
of the gas shell (see table \ref{table: model parameters}).  This is
nearly the same as the radiation mean free path which implies that
electron-ion energy exchange occurs as rapidly as electron radiative
cooling.  We find that $T_i \gg T_e$ immediately
behind the shock but the temperatures equalize $5 \times 10^{10}$~cm behind
the shock (Fig. \ref{fig: subtf.4.11}a) .  
Although this distance is about a factor of ten larger than $l_{ei}$, the 
length scale over which the cooling occurs is comparable to $l_{ei}$.  
Electron conduction and radiative heating raise the preshock electron
temperature by a small amount.

The radiation flux in this case is small and has a roughly exponential
profile (Fig. \ref{fig: subtf.4.11}b).  The radiation energy density is
much larger than the equilibrium value of $e_{r,eq} = a_r T_e^4$, as
expected for a subcritical shock (Sincell, \etal 1997).  Electron conduction 
smooths the gradient in the electron temperature profile, so the sharp peak 
in the radiation flux profile of the one-temperature gas (Sincell, \etal 1997) 
is rounded off.

The electron conduction flux at the shock front \begin{equation}
F_c = \rho \kappa_{ec} {\partial T_e \over \partial r}
\end{equation}
is many orders of magnitude larger than the radiation flux because of the
large gradient in $T_e$ at the shock front.  However, $l_{ec}$ is much smaller
than a radiation mean free path so electron conduction has a very
small
effect upon the temperature profile.

The shock wave becomes electron supercritical when $u_p \gtwid 10$~km/sec 
(Fig. \ref{fig: supertf.4.11}a).
The increased piston speed reduces $l_{ei}$ so that it is much smaller
than a radiation mean free path and the electron and ion gases 
are nearly decoupled.  The ions upstream from the shock are preheated by
the hot electron gas, but $T_e \gtwid T_i$ upstream from the shock because
of the large value of $l_{ei}$.  Similarly, the ion gas is cooled by
Coulomb collisions with the colder electron gas but $T_i \gtwid T_e$
downstream from the shock.  

The electron gas and the radiation field are in equilibrium from the shock
discontinuity to an optical depth of $\tau \sim 3$.  In this region the
gas temperature falls roughly as (see Sincell, \etal 1997)
\begin{equation}
T_e \propto \left( 1 + {3\sqrt{3} \over 4} |\tau-\tau_c| \right)^{1/3}
\end{equation}
where $\tau_c$ is the optical depth where the electron gas and the
radiation field fall out of equilibrium.

The ratio of the peak electron temperature to the post shock temperature is
nearly 1.5, as expected for a supercritical shock wave (Sincell, \etal 1997).  
Electronic heat
conduction does not reduce the amplitude of the spike because $l_{ec}$ is
much smaller than the length scale of the temperature spike.
Note that in Fig. \ref{fig: supertf.4.11}ab the $T_e$ and $F_c$ profiles
have
been offset from the $T_i$ and $F_r$ profiles so that the spikes at the
shock radius can be seen.

The radiation flux profile (Fig. \ref{fig: supertf.4.11}b)
is very similar to that of a supercritical wave in 
the one temperature fluid, again reflecting the weak coupling of the electron
and ion fluids and the small conduction length scale.  The peak value of
$F_c$ is again much larger than the peak of the radiation flux.  Electron
conduction is effective in keeping $T_e$  approximately constant near the
piston, whereas $T_i$ drops rapidly.

\subsection{$\Lambda_{ei} = 10^{-4}$~sec$^{-1}$, 
        $l_e \bar v_e = 10^{15}$~cm$^2/$~sec}
\label{subsec: small large}

Another example of an electron subcritical wave is displayed in Fig.
\ref{fig: subtf.4.15}.  In this case, we find that $l_{ec} \simeq l_{ei}
\simeq 10^9$~cm (table \ref{table: model parameters}).  The conduction
length is now comparable to the radiation mean free path, so electron
conduction is effective in preheating the upstream electron gas.  The
temperature profile of the preheated electrons falls exponentially on a
scale $l_{ec}$.  Conduction also removes the discontinuity in $T_e$ at the
shock radius (Fig. \ref{fig: subtf.4.15}a).

Both the conduction and radiation fluxes 
drop exponentially with distance from the shock (Fig. \ref{fig:
subtf.4.15}b).
The length scale for the $F_c$ profile is $l_{ec}$ so the profile is 
broader, and the peak flux is much lower, than in model 1.
The $F_r$ profile is largely unchanged when the conductivity is increased.  The
length scale for $F_r$ is set by the (constant) opacity and the final value of
$T_e$
behind the shock is very close to the value in model 1.
Conduction is only effective in regions where the $T_e$ gradient is large, so it
is not surprising that it has little effect upon the temperature behind the
shock.

The shock wave becomes electron supercritical
when $u_p = 14$~km/sec.  The $F_r$ and $T_e$ profiles for this model
are very similar to those in model 2 (Figs. \ref{fig: supertf.4.15}ab).  
The only effect that the increase in the
conductivity has is to reduce the $T_e$ spike to a small blip at the shock
radius. The width of this blip is on the order of $l_{ec}$.
Although radiation heats the electron gas to $T_e \gg T_i$, 
the electron-ion
exchange length is large and a discontinuity in $T_i$ remains
at the shock
radius.
The conduction flux is also smaller in this case.

\subsection{$\Lambda_{ei} = 10^{-2}$~sec$^{-1}$, 
        $l_e \bar v_e = 10^{11}$~cm$^2/$~sec}
\label{subsec: large small}
The length scale for the electron-ion energy exchange is reduced by two orders
of magnitude in these models.  We find that $T_e = T_i$ on virtually all length
scales and when plotted as a function of radius they are nearly indistinguishable
(Fig. \ref{fig: subtf.2.11}a).  However, if we plot $T_e$ and $T_i$ as a
function
of grid point (Fig. \ref{fig: subtf.2.11}b) we can see that $T_i > T_e$ in
the
shock front.  
The two temperatures rapidly equilibrate after the shock and the width of the 
spike in $T_i$ at the shock front is roughly $l_{ei}$.
This demonstrates the unique power of the adaptive grid to 
resolve 
physical quantities on multiple length scales.

The shock wave is subcritical when $u_p = 4$~km/sec and $F_r$ has the familiar
exponential profile.  The $F_c$ profile is strongly peaked at the shock radius
but it also has a broad peak behind the shock.  This is caused by the strong
electron-ion coupling which forces $T_e$ to follow $T_i$ and introduces
curvature into the $T_e$ profile.

The structure of the two-temperature shock wave in the
supercritical case (Fig. \ref{fig: supertf.2.11}abc) is almost
identical to a supercritical shock in the one-temperature approximation 
(paper I).  This
is not surprising because the small length scale for energy exchange implies
that the two fluids are completely coupled.  

The only departures from the 
one-temperature approximation occur in the shock front, where $T_e \ltwid
T_i$.  The reduction in the electron temperature reduces the flux from the
supercritical two-temperature shock by a few percent relative to the 
one-temperature fluid (Sincell, \etal 1997).
The electron conduction flux has a very large peak, caused by the spike in
$T_e$, but it has a negligible effect upon the temperature distribution because
$l_{ec}$ is small.

\subsection{$\Lambda_{ei} = 10^{-2}$~sec$^{-1}$, 
        $l_e \bar v_e = 10^{15}$~cm$^2/$~sec}
\label{subsec: large large}

Increasing the electron conductivity in the strongly coupled fluid does not
have any pronounced effects upon the structure of the shock wave (Figs.
\ref{fig: subtf.2.15}abc and \ref{fig: supertf.2.15}abc).  The enhanced 
conduction makes the post shock
gas in the subcritical flow nearly isothermal and there is a small amount
of preheating of the electron gas (Fig. \ref{fig: subtf.2.15}b).
We also see that the $F_c$ profile is slightly broadened ahead of the shock.

Electron conduction reduces the peak of $T_e$ in the supercritical case.  The
radiation flux is therefore smaller in this case.

\subsection{Critical Energy Exchange Rate}
\label{subsec: critical rate}

We find that no supercritical shock develops at any speed when $\tau_{ei}
\gtwid \tau_{ei,c} = 2 \times 10^5$~seconds because the low rate of 
electron-ion
energy exchange reduces the gradient in the electron energy density, limiting
the radiative flux.  The radiation flux from the 
shocked gas is given approximately by the diffusion equation
\begin{equation}
F_r \simeq {16 \sigma T^3 \over \rho \kappa} {dT \over dr}
\end{equation}
because the gas is optically thick.  The length scale for heating the electron
gas by electron-ion energy exchange is $l_{ei}$ so $F_r \ltwid F_c = 
\sigma T_c^4$ when
\begin{equation}
\tau_{ei} \gtwid \tau_{ei,c} = {16 \over 3} {T \over T_c} 
{1 \over \rho \kappa D}
\end{equation}
where $T_c$ and $F_c$ are the critical temperature and radiation flux (see
Sincell, \etal (1997)).  Assuming that $T = T_c$ and $D = 10$~km/sec, we find
$\tau_{ei,c} = 2 \times 10^5$~seconds.

\section{Conclusions}
\label{sec: conclusions}

We have solved the time-dependent spherically-symmetric equations of
radiation hydrodynamics on an adaptive grid for a fully ionized gas.  We
treat the gas as a two-temperature fluid and neglect recombination.
The time scale for electron-ion energy exchange, $\tau_{ei}$, and the
electron conduction coefficient, $\kappa_{ec}$, are assumed to be constant
free parameters of the problem.  The gas opacity is constant and purely
absorptive.

In this paper, the shock wave is created by moving a supersonic piston
into a constant density shell of cold, but fully ionized, gas.  The shock
propagates into the gas at a speed $D = u_p /(1-\eta_{+})$, where
$\eta_{+}$ is the compression ratio just downstream of the shock front,
and we find that the structure of the shock becomes steady in the shock
frame after a time short compared to the flow time.  The temperature of
the shocked gas is approximately $T_1 \simeq 27 u_{p,5}^2$~K, where
$u_{p,5}$ is the piston speed in $10^5$~cm/sec. 

Shocks in a two-temperature gas can be grouped into four general
categories depending on the piston speed and the length scale for
electron-ion energy exchange, $l_{ei} = \tau_{ei} D$.

For our model problem, the shock wave is electron subcritical when
$l_{ei} \gtwid 10^4$~km and $u_p \ltwid 9$~km/sec.  We find that $T_i >
T_e$ behind the electron subcritical shock because of the slow transfer of
energy from ions to the electrons.  There is no preheating of either
electrons or ions.  The shock becomes electron supercritical when $u_p
\gtwid 10$~km/sec.  In this case, preheating of the electron gas raises
the pre-shock $T_e$ to be equal to the post shock value.  However, the slow
transfer of energy from the electrons to the ions prevents preheating of
the ion gas.  Thus, $T_i > T_e$ behind the shock and $T_i < T_e$ ahead of
the shock.

We find that an electron supercritical shock cannot form when $\tau_{ei}
\gtwid 2 \times 10^5$~sec.  The transfer of energy is so slow that $T_e$
remains too low to preheat the electron gas. 

The electron and ion temperatures are nearly equal at all points in the
flow when $l_{ei} \ltwid 10^{3}$~km because the electron-ion energy
exchange is very efficient. For $u_p \ltwid 9$~km/sec, the shock wave is
subcritical and for larger $u_p$ it becomes supercritical. The structure
of the shock wave is nearly identical to that of a single temperature
fluid, as described in paper I. 

Electronic conduction smooths the $T_e$ profile on lengths $l_{ec} \sim
\kappa_{ec} / D$.  The radiation flux is roughly proportional to the
gradient in $T_e$, so larger conduction coefficients tend to reduce the
radiation flux.

\section{Figure and Table Captions}
\bigskip
Table \ref{table: model parameters}.  Parameters and length scales for the
models presented in this paper.  Symbols are defined in the text.

\begin{figure}
\plotone{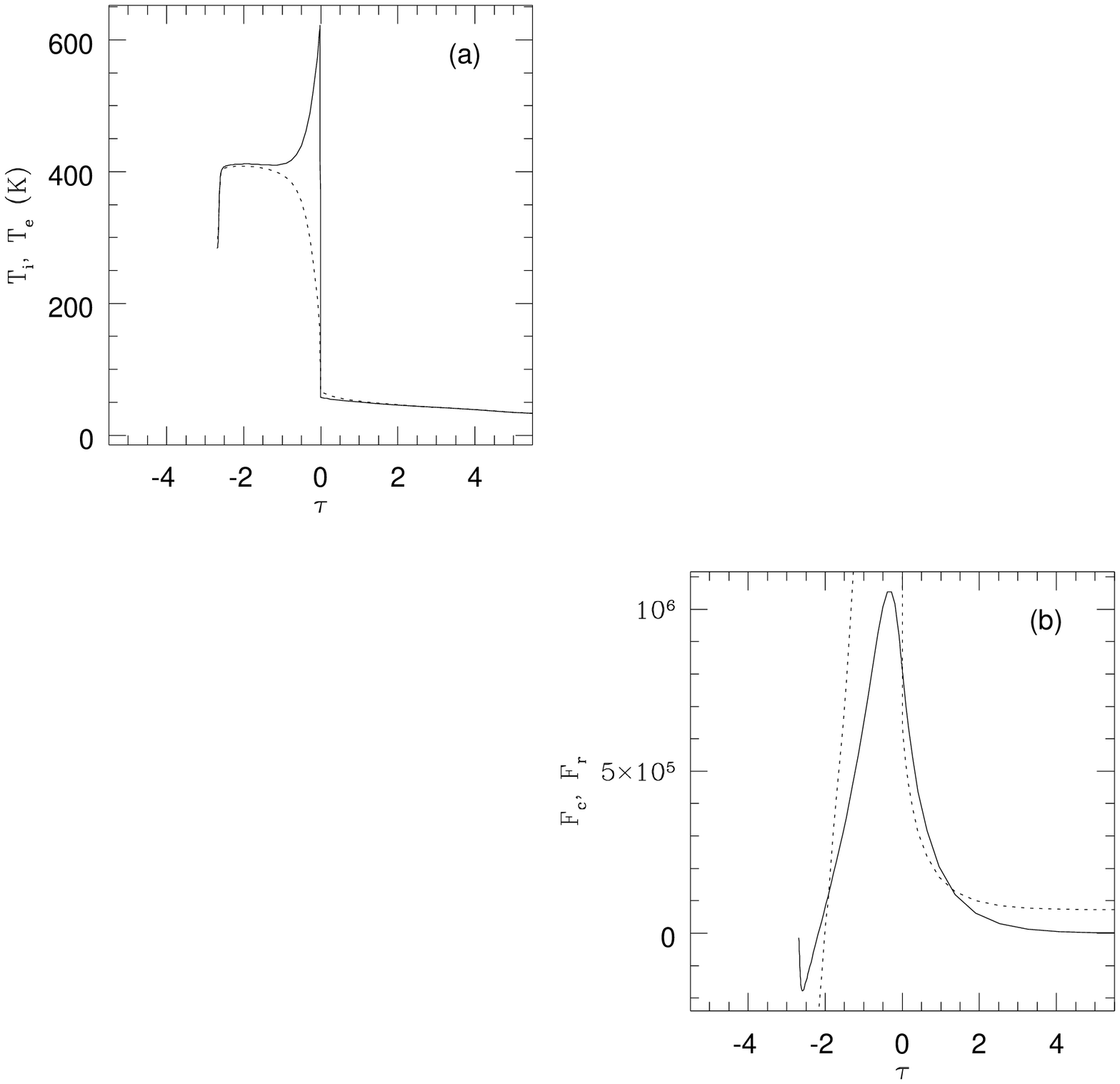}
\caption{\label{fig: subtf.4.11}   }
\end{figure}

Figure \ref{fig: subtf.4.11}.  The electron (dashed) and ion (solid) 
temperatures for model 1 
are plotted as a function of optical depth in (a). 
The corresponding radiation (solid) and electron conduction (dashed) fluxes
are in (b).  The peak of the electron conduction flux is much larger than
the scale of this plot.

\begin{figure}
\plotone{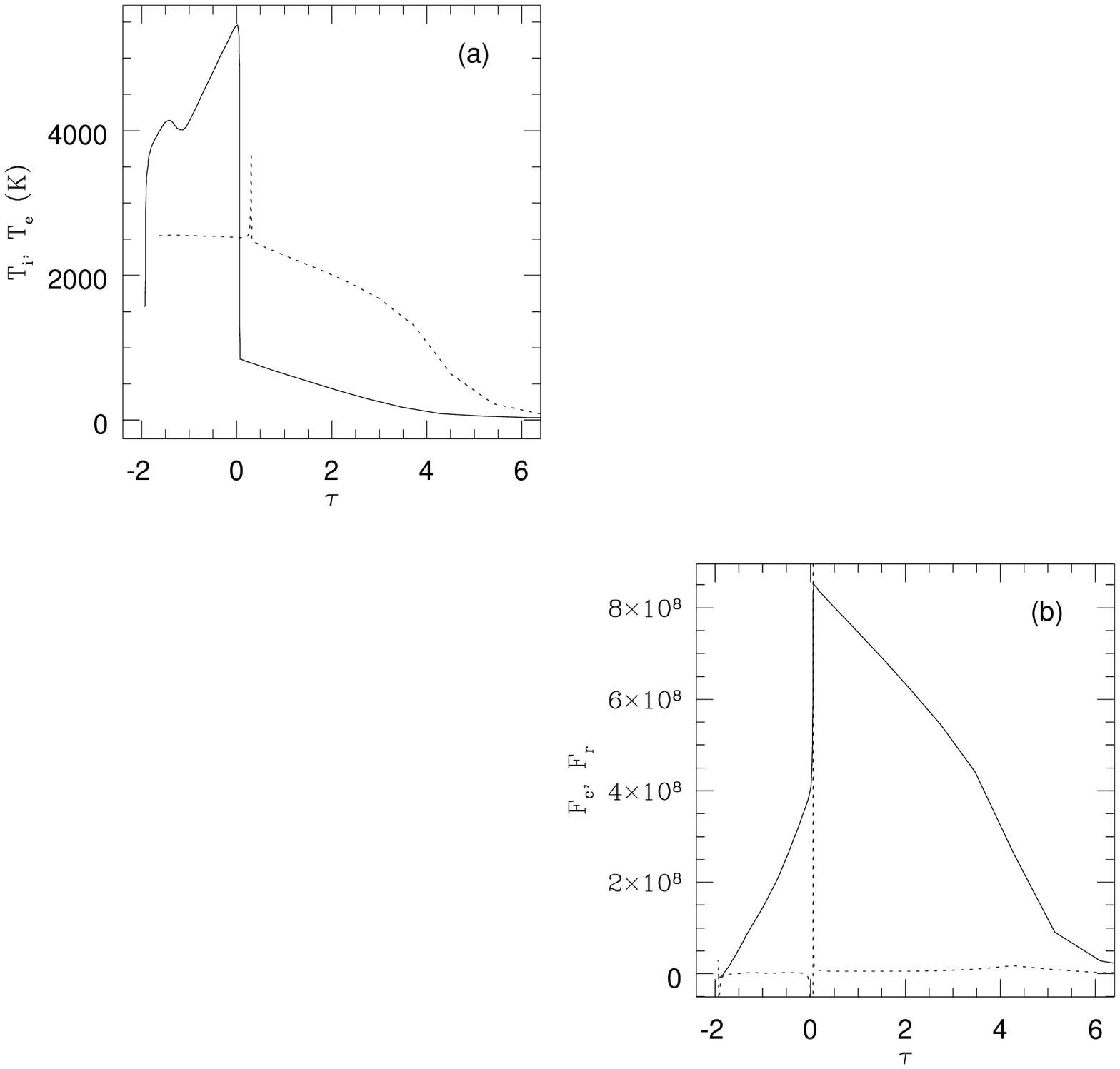}
\caption{\label{fig: supertf.4.11}   }
\end{figure}

Figure \ref{fig: supertf.4.11}.  Model 2. Note that the electron temperature
and the conduction flux profiles are displaced by a small amount to show the
spike at the shock radius. 

\begin{figure}
\plotone{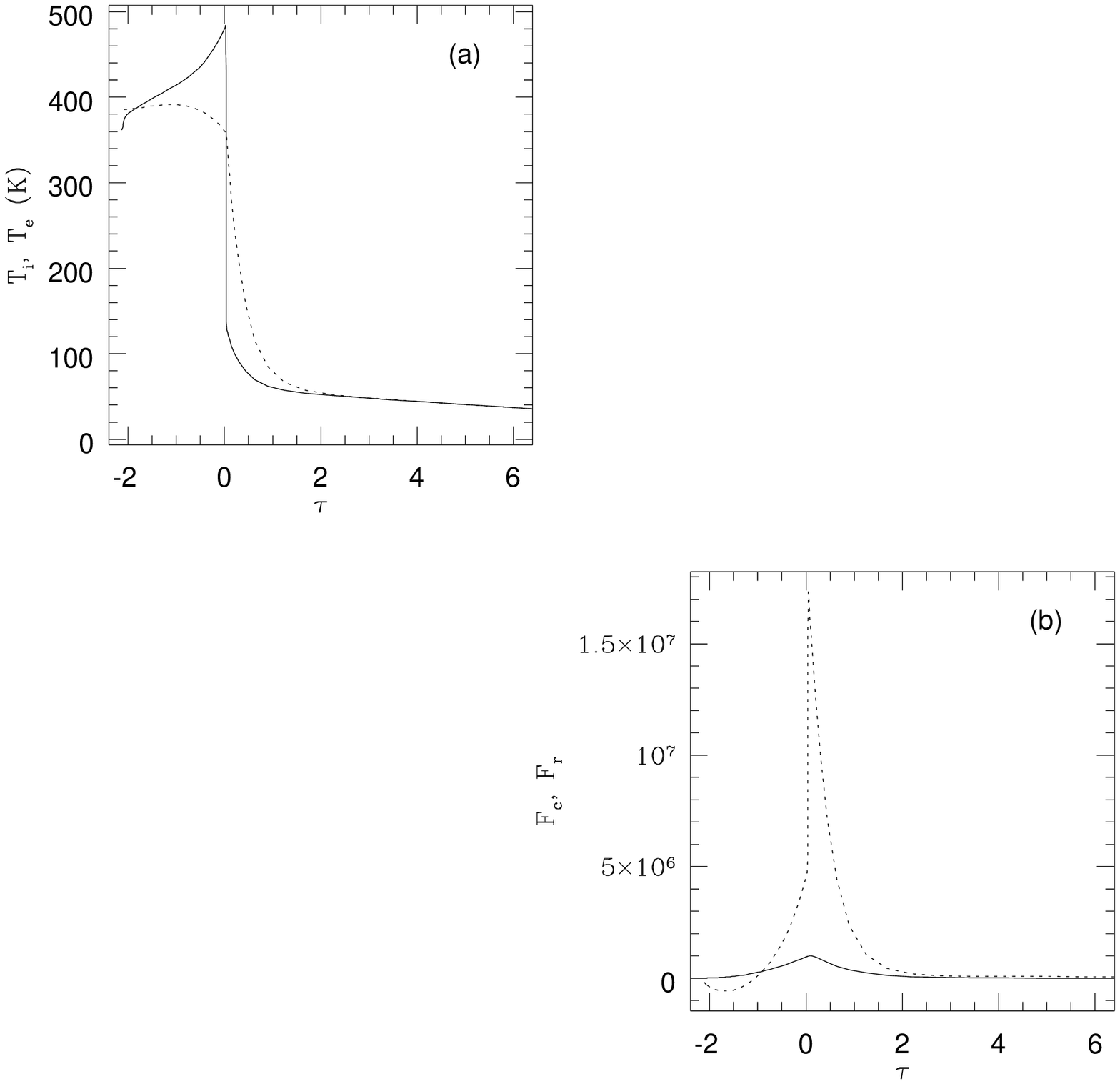}
\caption{\label{fig: subtf.4.15}   }
\end{figure}

Figure \ref{fig: subtf.4.15}.  Model 3 

\begin{figure}
\plotone{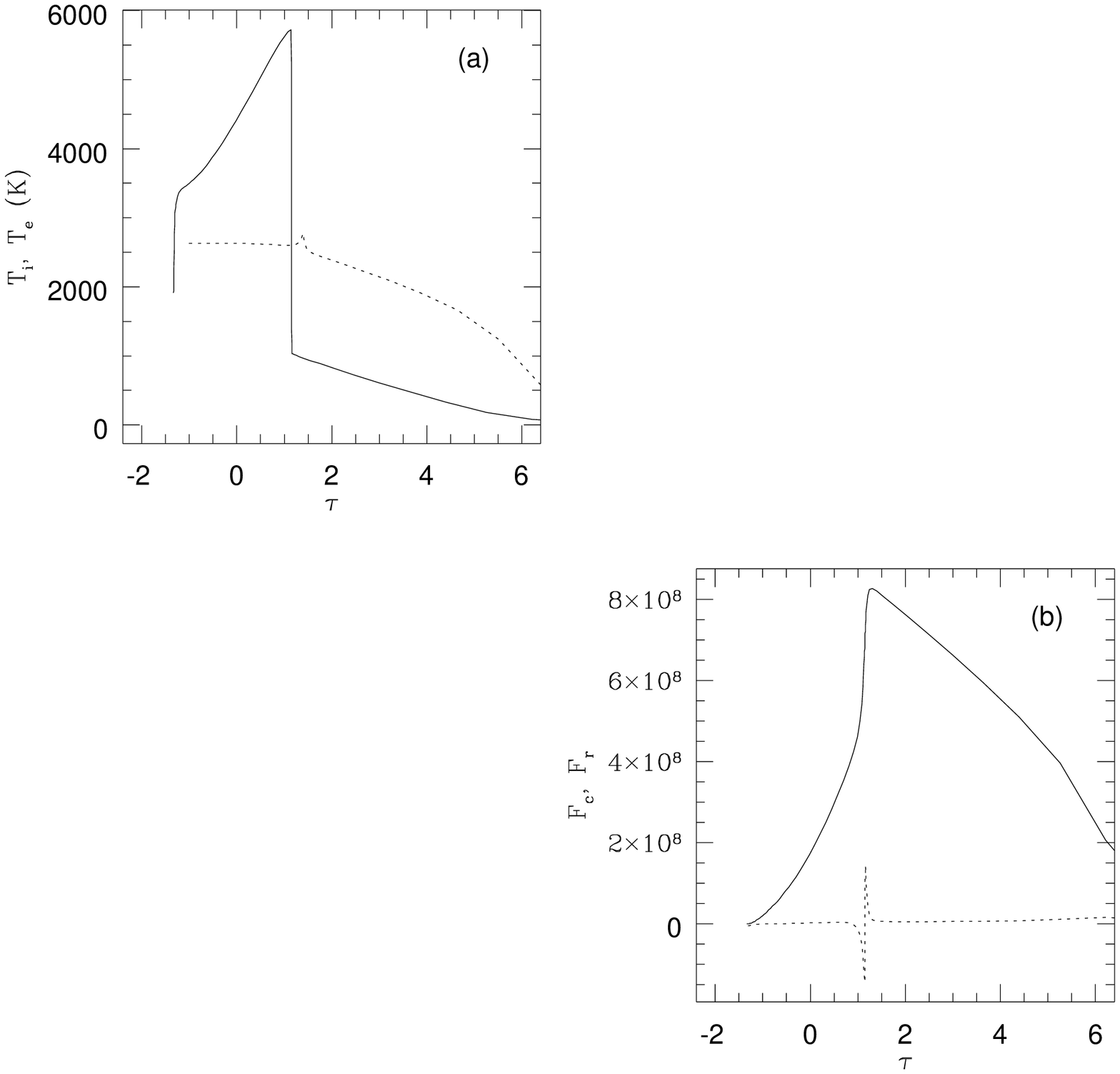}
\caption{\label{fig: supertf.4.15}   }
\end{figure}

Figure \ref{fig: supertf.4.15}.  Model 4.  The electron temperature has
been displaced by a small amount to show the temperature spike at the shock 
radius. 

\begin{figure}
\plotone{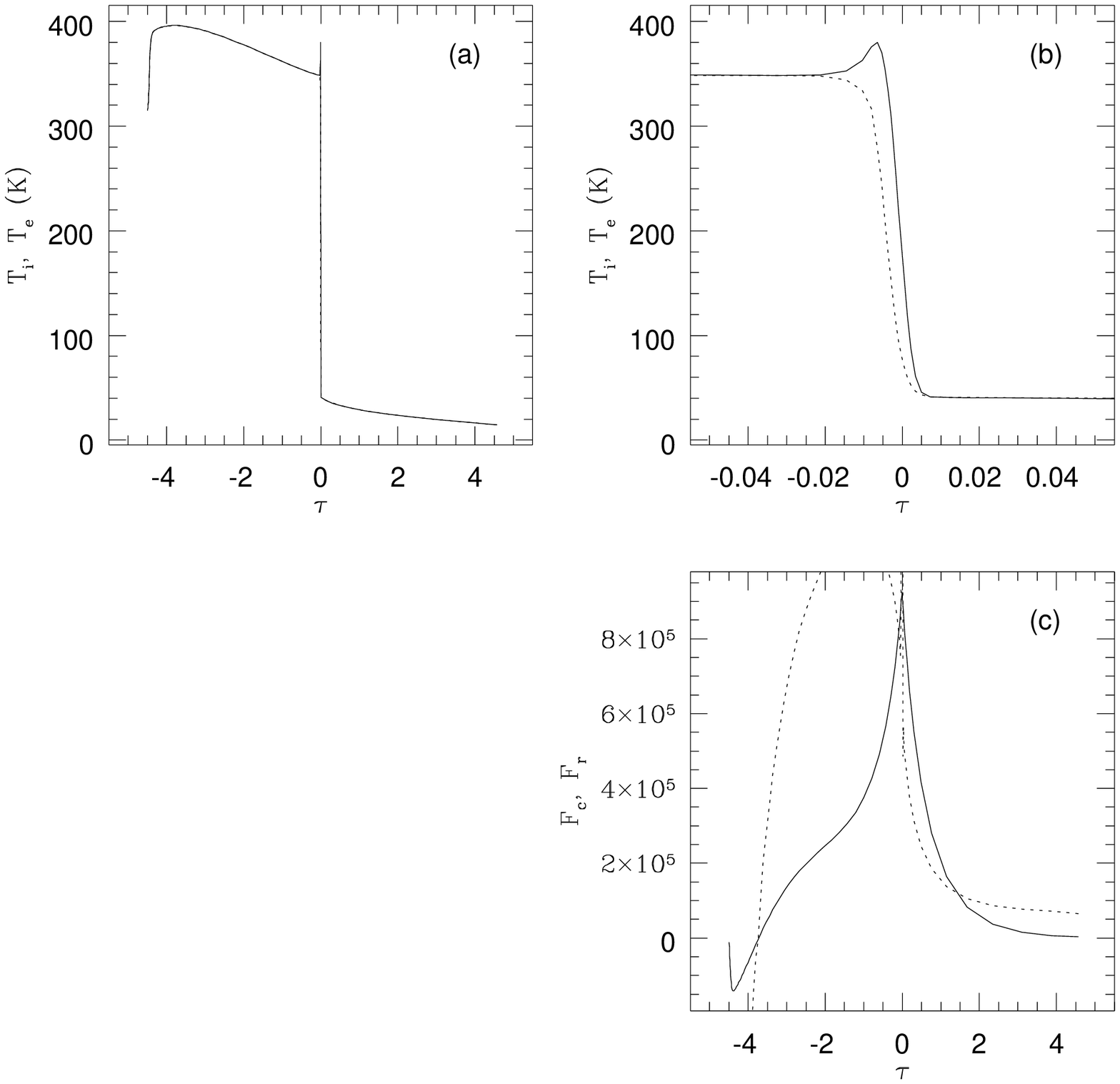}
\caption{\label{fig: subtf.2.11}   }
\end{figure}

Figure \ref{fig: subtf.2.11}.  The electron and ion temperatures for model 5 
are plotted as a function of optical depth in (a) and as a function of grid 
index in (b). 
The corresponding radiation and electron conduction fluxes
are in (c).  The peak electron conduction flux is larger than the scale of this
plot.

\begin{figure}
\plotone{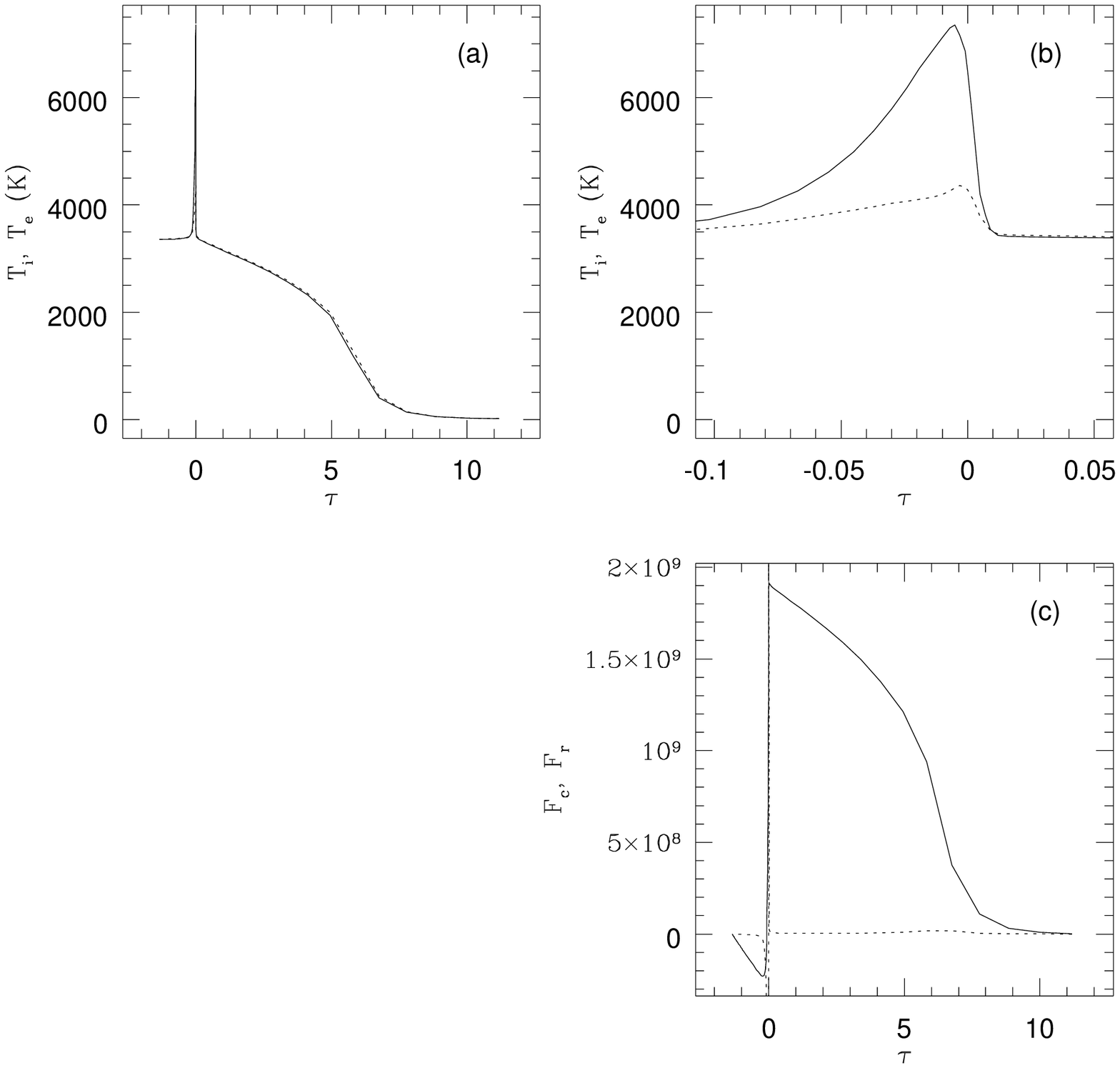}
\caption{\label{fig: supertf.2.11}   }
\end{figure}

Figure \ref{fig: supertf.2.11}.  Model 6. The peak of the electron conduction
flux is larger than the scale of this plot. 

\begin{figure}
\plotone{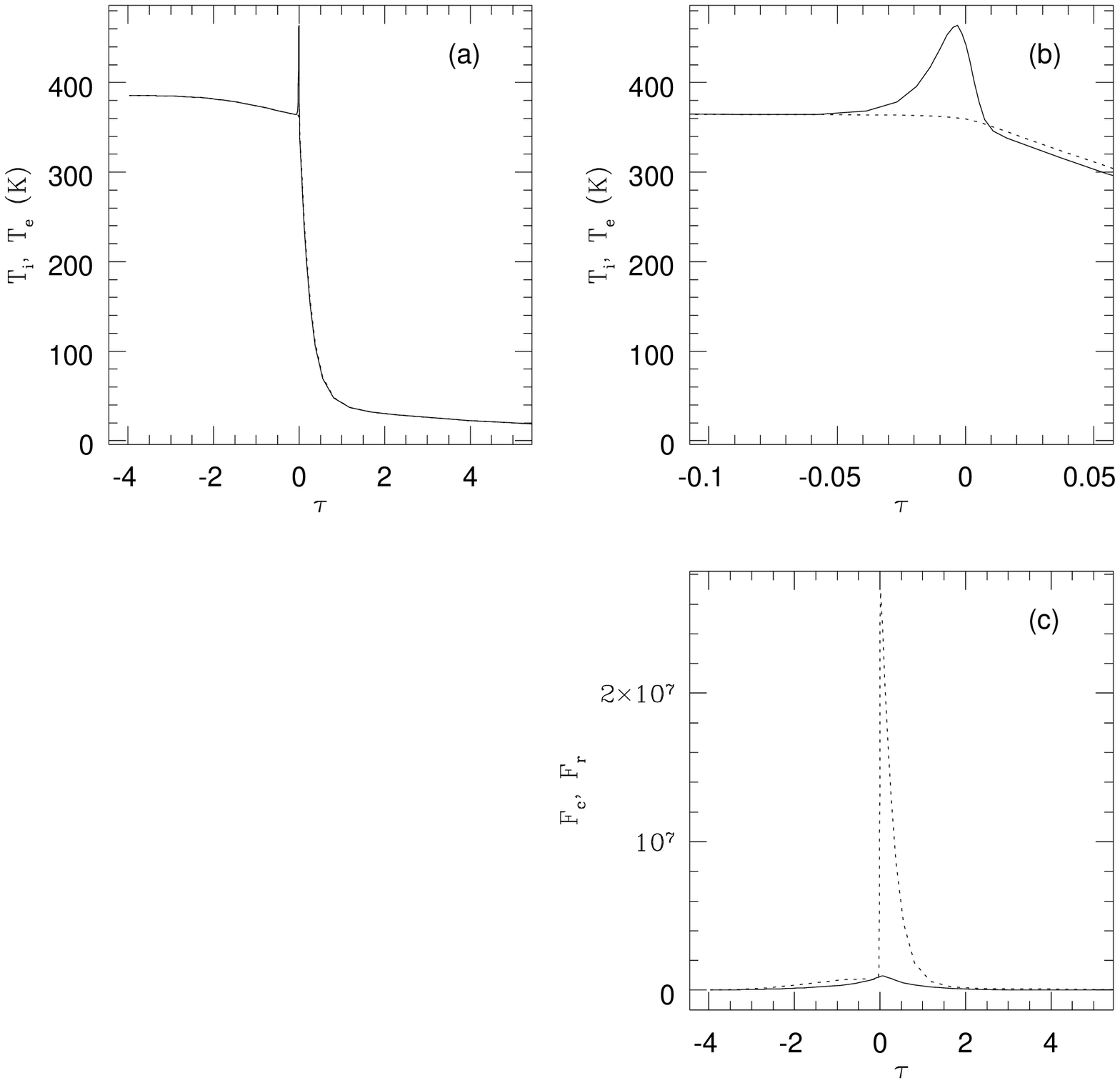}
\caption{\label{fig: subtf.2.15}   }
\end{figure}

Figure \ref{fig: subtf.2.15}.  Model 7 

\begin{figure}
\plotone{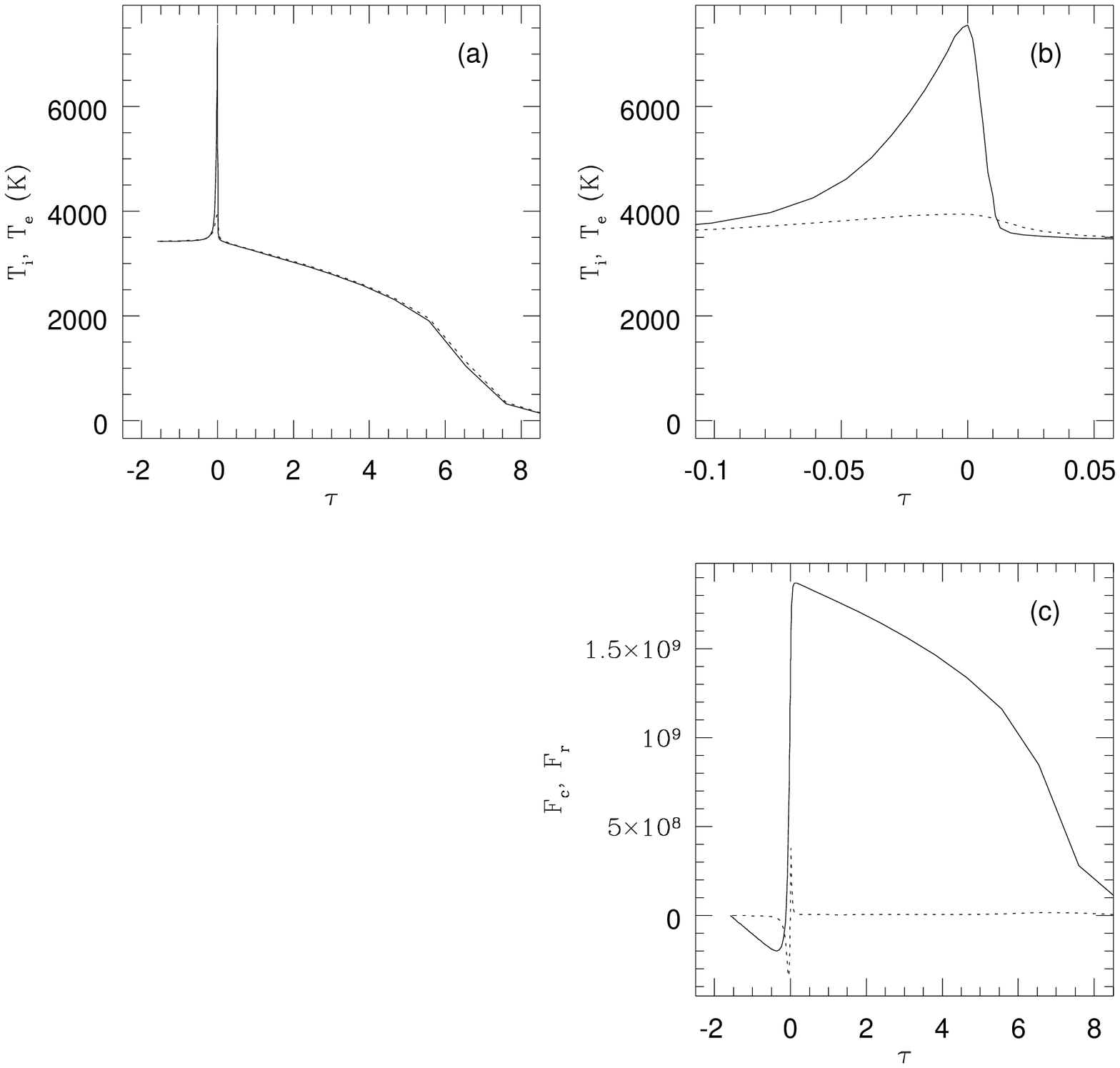}
\caption{\label{fig: supertf.2.15}   }
\end{figure}

Figure \ref{fig: supertf.2.15}.  Model 8 

\vfil\eject
\begin{table}
\begin{tabular}{||c|c|c|c|c|c||} \hline
Model & $\Lambda_{ei} (\mbox{sec}^{-1})$ & $(l \bar v)_c (\mbox{cm}^2/
\mbox{sec})$ 
& $u_p (\mbox{km/sec})$ & $l_{ei}$~cm & $l_{ec}$~cm \\ \hline
1 &  $10^{-4}$ & $10^{11}$ & 4 & $4 \times 10^{9}$ & $3 \times 10^{5}$  \\ 
2 & & & 14 & $ 1 \times 10^{10}$ & $9 \times 10^{4}$  \\ \hline
3 & $10^{-4}$ & $10^{15}$ & 4 & $4 \times 10^{9}$ & $3 \times 10^{9}$  \\ 
4 & & & 14 & $ 1 \times 10^{10}$ & $9 \times 10^{8}$  \\ \hline
5 & $10^{-2}$ & $10^{11}$ & 4 & $4 \times 10^{7}$ & $3 \times 10^{5}$  \\ 
6 & & & 14 & $ 1 \times 10^{8}$ & $9 \times 10^{4}$  \\ \hline
7 &  $10^{-2}$ & $10^{15}$ & 4 & $4 \times 10^{7}$ & $3 \times 10^{9}$  \\ 
8 & & & 14 & $ 1 \times 10^{8}$ & $9 \times 10^{8}$  \\ \hline
\end{tabular}
\caption{\label{table: model parameters}}
\end{table}

\begin{references}
\noindent Burger HL, Katz JI (1983)  The Eddington limit and supercritical
accretion II: Time-dependent calculations. Astroph. J. 265:393-401

\noindent Dorfi EA, Drury LO'C, (1987) Simple adaptive grids for 1D initial 
value problems. Physica D. 69:175

\noindent Gehmeyr M, Mihalas D, (1994) Adaptive grid radiation hydrodynamics 
with TITAN. Physica D 72:320-???

\noindent Mihalas D, Mihalas BW (1984) Foundations of Radiation
Hydrodynamics, Oxford University Press, Oxford

\noindent Tscharnuter WM, Winkler KH (1979) A method for computing
self gravitating
gas flows with radiation. Comput. Phys.
Commun. 18:171

\noindent Winkler KH, Norman ML, Mihalas D (1984) Adaptive mesh radiation
hydrodynamics--I. The radiation transport equation in a completely
adaptive coordinate system. J.Q.S.R.T. 31:473-478

\noindent Klein RI, Stockman HS, Chevalier RA (1983) Supercritical
time-dependent accretion onto compact objects I: Neutron stars. Ap.J.
237:912-930

\noindent Shafranov VD (1957), The structure of shock waves in a plasma.
Sov. Physics JETP 5:1183-1188

\noindent Sincell MW, Gehmeyr M, Mihalas D (1997), The Quasi-stationary
Structure of Radiating Shock Waves I: The One-temperature Fluid, to appear
in Shock Waves

\noindent Zel'dovich YB, Raizer YP (1967) Physics of Shock Waves
and High Temperature Hydrodynamic Phenomena, Academic Press, New York
                
\end{references}
\end{document}